\documentclass[letter,traditabstract]{aa} 
\usepackage{graphicx}
\usepackage{txfonts}
\usepackage{natbib}
\newcommand{\tablenotea}[1]{\parbox{8.8cm}{ \indent
\footnotesize{\textsc{Note.--}~#1}}}
\newcommand{\irpc}{Int.~Rev.~Phys.~Chem.}         %
\newcommand{\scactaa}{Spectrochim.~Acta A}        %
\newcommand{\molphys}{Mol.~Phys}                  %
\newcommand{\prsla}{Proc.~R.~Soc.~London, Ser.~A} %
\newcommand{\jmst}{J.~Mol.~Struct.}               %
\newcommand{\chemrev}{Chem.~Rev.}                 %

\begin{document}
\title{Astronomical identification of CN$^-$, the smallest observed molecular anion\thanks{Based on
observations carried out with the IRAM 30-meter telescope. IRAM is
supported by INSU/CNRS (France), MPG (Germany) and IGN (Spain).}}
\titlerunning{Astronomical identification of CN$^-$}
\authorrunning{Ag\'undez et al.}
\author{M. Ag\'undez\inst{1}, J. Cernicharo\inst{2}, M. Gu\'elin\inst{3}, C. Kahane\inst{4}, E.
Roueff\inst{1}, J. K\l os\inst{5}, F. J. Aoiz\inst{6}, F.
Lique\inst{7}, \\ N. Marcelino\inst{2}, J. R. Goicoechea\inst{2},
M. Gonz\'alez Garc\'ia\inst{8}, C. A. Gottlieb\inst{9} , M. C.
McCarthy\inst{9} and P. Thaddeus\inst{9}} \institute{LUTH,
Observatoire de Paris-Meudon, 5 Place Jules Janssen, 92190 Meudon,
France; \email{marcelino.agundez@obspm.fr} \and Departamento de
Astrof\'isica, Centro de Astrobiolog\'ia, CSIC-INTA, Ctra. de
Torrej\'on a Ajalvir km 4, 28850 Madrid, Spain \and Institut de
Radioastronomie Millim\'etrique, 300 rue de la Piscine, 38406
Saint Martin d'H\'eres, France \and Laboratoire d'Astrophysique de
l'Observatoire de Grenoble, 38041 Grenoble, France \and Department
of Chemistry and Biochemistry, University of Maryland, College
Park, MD 20742, USA \and Departamento de Qu\'imica F\'isica,
Facultad de Qu\'imica, Universidad Complutense, 28040 Madrid,
Spain \and LOMC – FRE 3102, CNRS – Universit\'e du Havre, 25 rue
Philippe Lebon, BP 540, 76058 Le Havre, France \and Instituto de
Radioastronom\'ia Milim\'etrica, Av Divina Pastora 7, Local 20,
18012, Granada, Spain \and Harvard-Smithsonian Center for
Astrophysics, 60 Garden Street, Cambridge, MA 02138, USA}

\date{Received; accepted}


\abstract
{We present the first astronomical detection of a diatomic
negative ion, the cyanide anion CN$^-$, as well as quantum
mechanical calculations of the excitation of this anion through
collisions with para--H$_2$. CN$^-$ is identified through the
observation of the $J$ = 2-1 and $J$ = 3-2 rotational transitions
in the C-star envelope IRC +10216 with the IRAM 30-m telescope.
The U-shaped line profiles indicate that CN$^-$, like the large
anion C$_6$H$^-$, is formed in the outer regions of the envelope.
Chemical and excitation model calculations suggest that this
species forms from the reaction of large carbon anions with N
atoms, rather than from the radiative attachment of an electron to
CN, as is the case for large molecular anions. The unexpectedly
large abundance derived for CN$^-$, 0.25 \% relative to CN, makes
likely its detection in other astronomical sources. A parallel
search for the small anion C$_2$H$^-$ remains so far unconclusive,
despite the previous tentative identification of the $J$ = 1-0
rotational transition. The abundance of C$_2$H$^-$ in IRC +10216
is found to be vanishingly small, $<$ 0.0014 \% relative to
C$_2$H.}
{}
{}
{}
{}

\keywords{astrochemistry --- line: identification --- molecular
processes --- stars: AGB and post-AGB --- circumstellar matter ---
stars: individual (IRC +10216)}

\maketitle
%

\section{Introduction}

The molecular anions detected so far in the interstellar and
circumstellar gas are all fairly heavy linear carbon chains with
three or more carbon atoms, and with neutral counterparts with
large electron affinities: C$_4$H$^-$, C$_6$H$^-$, C$_8$H$^-$,
C$_3$N$^-$, and C$_5$N$^-$
\citep{mcc06,cer07,bru07a,rem07,tha08,cer08}. The abundance of
these anions relative to the neutral counterparts increases with
size and with the electron affinity of the neutral molecule, as
expected for formation by radiative electron attachment
\citep{her08}. On inspection, however, that process fails to
explain the abundance of the shortest observed anions, in
particular C$_4$H$^-$ and C$_3$N$^-$. In IRC +10216, a carbon star
envelope where both anions are found, C$_3$N$^-$ has an
anion-to-neutral abundance ratio about 50 times higher than that
of C$_4$H$^-$, indicating that other formation processes may be at
work \citep{cer07,tha08,agu09,cor09}. Studying the astronomical
abundance of even shorter anions, in particular C$_2$H$^-$ and
CN$^-$, whose formation by radiative electron attachment is very
slow, should help answer this question.

In this Letter we describe the identification in IRC +10216 of
CN$^-$ and the results on a parallel search for C$_2$H$^-$. We
also present quantum mechanical calculations of the collisional
excitation of CN$^-$ by para--H$_2$, using the calculated rate
coefficients to model the observed lines. The chemistry of CN$^-$
in space is also briefly discussed.

\section{Observations and identification of CN$^-$}

The C$_2$H$^-$ and CN$^-$ anions are closed-shell molecules whose
rotational spectrum has been recently measured in the laboratory
\citep{bru07b,got07,ama08}. Their electric dipole moments are 3.1
and 0.65 Debye respectively \citep{bru07b,bot95}.

\begin{figure}
\includegraphics[angle=0,scale=.545]{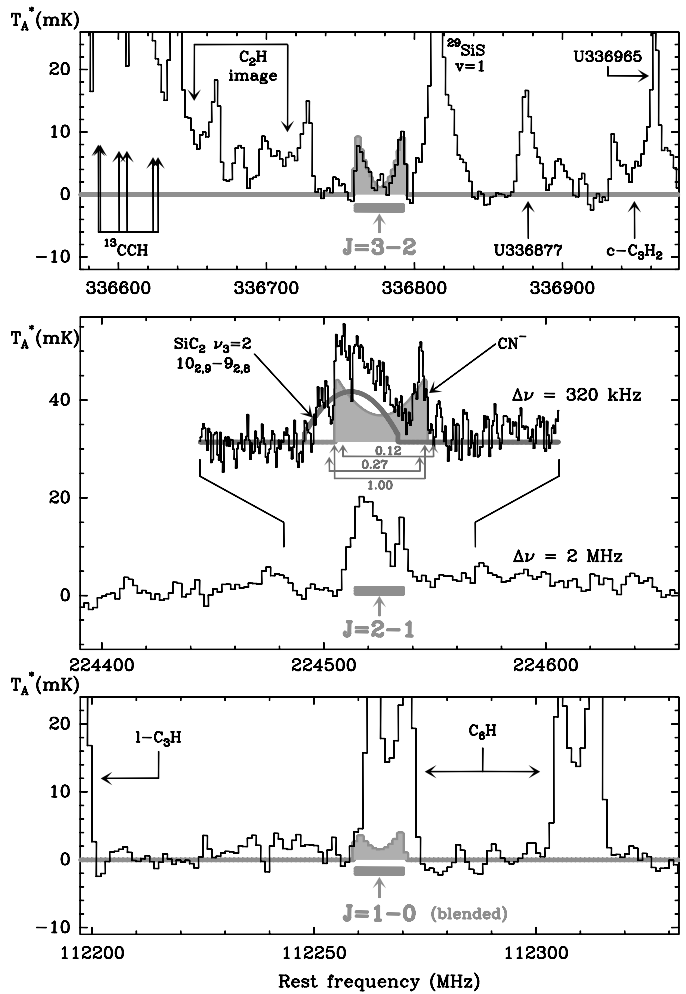}
\caption{Spectra of IRC +10216 covering the $J$ = 1-0 to $J$ = 3-2
transitions of CN$^-$. Grey horizontal boxes mark their expected
positions based on the laboratory frequencies and a linewidth of
29 km s$^{-1}$. Shaded areas show the fits to the line profiles
obtained with the \emph{CLASS} method \emph{shell}. The high
spectral resolution spectrum of the $J$ = 2-1 line shows the
expected position of the different hyperfine components with their
relative intrinsic strengths. The intensity scale is expressed as
$T_A^*$, antenna temperature corrected for atmospheric absorption
and antenna ohmic and spillover losses. To transform into main
beam brightness temperature ($T_{\rm MB}$) in this Figure and in
Table~\ref{table-lineparameters} divide by 0.78, 0.65, and 0.44 at
112, 224, and 336 GHz, respectively.} \label{fig-cnminus-lines}
\vspace{-0.1cm}
\end{figure}

The present astronomical observations were carried out towards IRC
+10216 with the IRAM 30-m telescope on Pico Veleta (Spain). The
$J$ = 1-0 rotational transition of CN$^-$ at 112.3 GHz was
observed before 2009 in the course of a $\lambda$ 3 mm spectral
survey (Cernicharo et al. in preparation). In IRC +10216's
spectrum, this line is severely blended with a strong component of
the $^2\Pi_{3/2}$ $J$ = 81/2-79/2 transition of C$_6$H (see
Fig.~\ref{fig-cnminus-lines}). The $J$ = 2-1 and $J$ = 3-2
rotational transitions of CN$^-$, at 224.5 and 336.8 GHz
respectively, were observed between January and April 2010 with
the new dual polarization EMIR receivers operating in single side
band mode. The rejection of the image side band was 13$-$15 dB at
224 GHz and 20$-$30 dB at 336 GHz, depending on the polarization,
as measured with strong lines. The local oscillator was shifted in
frequency to identify possible contamination from the image side
band. The backends were two autocorrelators with 2 MHz and 320 kHz
channel spacings, respectively. The pointing and focus of the
telescope were checked every 1-2 hours on Mars and on the nearby
quasar OJ 287. To obtain flat baselines, the secondary mirror was
wobbled by 180$''$ at a rate of 0.5 Hz. The zenith sky opacity at
225 GHz was typically $\sim$ 0.1, resulting in system temperatures
of 140 K at 224 GHz and of 800 K at 336 GHz. The total integration
time per polarization was 3 h at 224 GHz and 9.5 h at 336 GHz,
yielding a rms noise of $T_A^*$ $\sim$ 2 mK per 2 MHz channel at
both frequencies, after averaging both polarizations.

Following our initial detection in IRC +10216 of a $T_A^*$ $\sim$
3 mK line at the frequency of the $J$ = 1-0 transition of
C$_2$H$^-$ \citep{cer08}, we searched from January to April 2010
for the $J$ = 2-1 transition at 166.5 GHz. The line was not
detected with a $T_A^*$ rms noise level of 0.6 mK per 2 MHz
channel, casting doubt on the tentative identification of the
C$_2$H$^-$ $J$ = 1-0 line.

\begin{table}
\caption{Observed line parameters of CN$^-$}
\label{table-lineparameters} \centering
\begin{tabular}{crllc}
\hline \hline
\multicolumn{1}{c}{}           & \multicolumn{1}{c}{$\nu_{\rm 0}$$^a$} & \multicolumn{1}{c}{$\nu_{\rm obs}$} & \multicolumn{1}{c}{v$_{\rm exp}$$^b$} & \multicolumn{1}{c}{$\int$$T_A^*$$d$v} \\
\multicolumn{1}{c}{Transition} & \multicolumn{1}{c}{(MHz)}         & \multicolumn{1}{c}{(MHz)}           & \multicolumn{1}{c}{(km s$^{-1}$)} & \multicolumn{1}{c}{(K km s$^{-1}$)} \\
\hline
$J$ = 1-0 & 112264.8  & 112264.8$^c$  & 14.5$^c$  & $\sim$ 0.07(3)$^d$ \\
$J$ = 2-1 & 224525.1  & 224525.4(5)   & 14.5$^c$  & 0.23(7)$^e$ \\
$J$ = 3-2 & 336776.4  & 336777.0(12)  & 15.0(10)  & 0.13(2) \\
\hline
\end{tabular}
\tablenotea{Number in parentheses are 1$\sigma$ uncertainties in
units of the last digits. $^a$ Frequencies derived from the
rotational constants reported by \citet{ama08}. $^b$ v$_{\rm exp}$
is the half width at zero level. $^c$ Fixed value. $^d$ Highly
uncertain estimate. Line severely blended with a strong C$_6$H
line. $^e$ Line blended with a SiC$_2$ $\nu_3$=2 line.}
\end{table}

The CN$^-$ observed lines are shown in
Fig.~\ref{fig-cnminus-lines} and the derived line parameters are
given in Table~\ref{table-lineparameters}. The $J$ = 3-2
transition of CN$^-$ is shown in the top panel of
Fig.~\ref{fig-cnminus-lines}. It appears as a U-shaped line with
the expected half width (v$_{\rm exp}$ = 15 $\pm$ 1 km s$^{-1}$)
that agrees in frequency to within 0.6 MHz with that of the CN$^-$
transition. The $J$ = 2-1 transition of CN$^-$, shown in the
middle panel of Fig.~\ref{fig-cnminus-lines} with a spectral
resolution of 2 MHz and of 320 kHz (2.7 and 0.4 km s$^{-1}$
respectively), coincides with a broad spectral feature with a
complex shape that is unusual for IRC +10216, since it is neither
U-shaped, flat-topped or parabolic. It is best explained as a
blend, as shown in Fig.~\ref{fig-cnminus-lines}, that can be well
fitted with two components, one U-shaped with a half width v$_{\rm
exp}$ of 14.5 km s$^{-1}$ centered at the frequency of the $J$ =
2-1 transition of CN$^-$ (see Table~\ref{table-lineparameters}),
the other with a parabolic profile, a half width v$_{\rm exp}$ of
15 $\pm$ 3 km s$^{-1}$, and a rest frequency of 224518.3 $\pm$ 1.5
MHz that is close to that of the 10$_{2,9}$-9$_{2,8}$ rotational
transition of SiC$_2$ in the $\nu_3$=2 vibrational state (224519.7
MHz; \citealt{izu94}). Since other $\nu_3$=2 lines of SiC$_2$ with
similar intrinsic strengths have similar shapes, half widths
(v$_{\rm exp}$ = 8-15 km s$^{-1}$), and intensities ($T_A^*$
$\sim$ 20 mK) in our $\lambda$ 0.9 mm data (Kahane et al. in
preparation) as our fitted parabolic component, there is little
doubt that this component comes from SiC$_2$. We note that the
CN$^-$ $J$ = 2-1 transition has several hyperfine components due
to the nitrogen quadrupole, which can be grouped into three blocks
lying at 224523.9, 224525.1, and 224527.2 MHz, with relative line
strengths of 0.27, 1, and 0.12, respectively \citep{got07}. Due to
the severe blending with the SiC$_2$ $\nu_3$=2 and to the limited
sensitivity of the astronomical observations, only the strongest
hyperfine component is clearly visible in the spectrum of IRC
+10216, while the middle strength component is hidden between the
two stronger fitted lines (see Fig.~\ref{fig-cnminus-lines}), and
the weakest hyperfine component lies below the noise level of the
spectrum. Finally, the bottom panel of
Fig.~\ref{fig-cnminus-lines} shows the spectrum covering the
CN$^-$ $J$ = 1-0 transition, which is heavily blended with a
strong line of C$_6$H. The limited spectral resolution (1 MHz) and
the broadening of this CN$^-$ line by the hyperfine structure
(there are three components separated by 1-2 MHz; \citealt{got07})
makes it difficult to determine the relative contributions of
C$_6$H and CN$^-$ to the observed line.

There are no good candidates other than CN$^-$ for the carrier of
the 336777.0 MHz line. The only plausible molecule with a
transition within 2 MHz of the observed frequency, according to
the line catalogs of J. Cernicharo, CDMS \citep{mul05}, and JPL
\citep{pic98}, is $^{13}$CCH, whose $N_{J,F_1,F}$ =
4$_{7/2,4,7/2}$-3$_{7/2,4,7/2}$ transition lies at 336775.7 MHz.
This molecule, however, is ruled out since the nearby
4$_{7/2,4,9/2}$-3$_{7/2,4,9/2}$ transition at 336756.2 MHz, with a
slightly higher intrinsic strength, is not present in our data
(see Fig.~\ref{fig-cnminus-lines}). Since no other plausible
candidate can be found for the 224525.4 MHz line and since
unidentified lines of that intensity are rare in IRC +10216 at
these frequencies, we conclude that we have almost certainly
detected CN$^-$. Confirmation of this identification would be
highly desirable, but may not be easy to obtain. The next two
rotational transitions of CN$^-$, at 449 and 561 GHz, cannot be
observed from ground owing to high atmospheric opacity, and still
higher $J$ transitions may be too weak to detect in a cool source
such as the outer envelope of IRC +10216.

The $J$ = 3-2 line of CN$^-$, which appears free of contamination
by background lines, has a pronounced U-shaped profile, which for
a spherical expanding envelope indicates that the emission is more
extended than the half-power beam of the telescope (7'' at 336
GHz). Thus CN$^-$ appears confined to the same outer envelope of
IRC +10216, as that of other molecular anions observed in this
source (e.g. \citealt{cer07,tha08,cer08}). A column density of 5
$\times$ 10$^{12}$ cm$^{-2}$ and a rotation temperature of 16 K
were derived from a rotational diagram constructed with the
velocity integrated intensities of the $J$ = 2-1 and 3-2 lines
given in Table~\ref{table-lineparameters}, on the assumption of a
uniform source with a radius of 20'' which is typical of molecules
distributed in the outer shell. The rotation temperature is
consistent with CN$^-$ emission from the cool outer envelope. With
a column density of the CN radical of 2 $\times$ 10$^{15}$
cm$^{-2}$, derived from several hyperfine components of the $N$ =
1-0 and $N$ = 3-2 transitions, we estimate a CN$^-$/CN abundance
ratio of 0.25 \%, which is comparable to the C$_3$N$^-$/C$_3$N
ratio in this source (0.52 \%; \citealt{tha08}).

From the upper limit of the $J$ = 2-1 line of C$_2$H$^-$, a
3$\sigma$ column density of $<$ 7 $\times$ 10$^{10}$ cm$^{-2}$ was
derived on the assumption of a source with a radius of 20'' and a
rotation temperature of 20 K. The estimated C$_2$H$^-$/C$_2$H
abundance ratio ($<$ 0.0014 \%) is at least 5 times smaller than
the already small C$_4$H$^-$/C$_4$H ratio \citep{agu09}.

\section{Modeling and discussion}

\begin{figure}
\includegraphics[angle=-90,scale=.365]{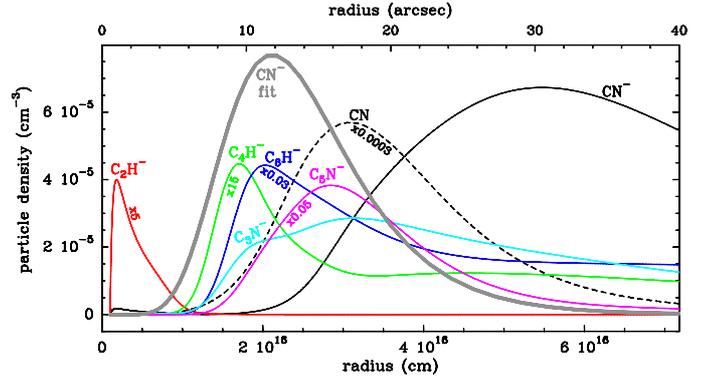}
\caption{Abundance distribution derived for CN$^-$ in the envelope
of IRC +10216 (thick grey line labeled as "CN$^-$ fit"), as it
reproduces the CN$^-$ observed line profiles (see
Fig.~\ref{fig-cnminus-lineprofiles}). Also shown are the
abundances of CN$^-$, CN, and other molecular anions as calculated
with the chemical model (multiplied by 0.0003, 5, 15, 0.03, and
0.05 for CN, C$_2$H$^-$, C$_4$H$^-$, C$_6$H$^-$, and C$_5$N$^-$,
respectively). The abundances are expressed as number of molecules
per cubic centimeter. The angular distance is given in the top
axis for an assumed distance to IRC +10216 of 120 pc.}
\label{fig-chemistry}
\end{figure}

\begin{figure}
\includegraphics[angle=-90,scale=.385]{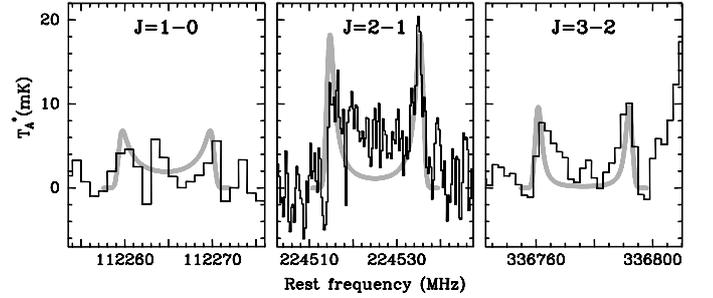}
\caption{Line profiles calculated with the LVG model (thick grey
lines) using the compact CN$^-$ abundance distribution (thick grey
line in Fig~\ref{fig-chemistry}) are compared with the observed
CN$^-$ lines (black histograms). Fits to the C$_6$H and SiC$_2$
$\nu_3$=2 lines have been subtracted in the $J$ = 1-0 and 2-1
observed spectra. The $J$ = 1-0 line profile is very uncertain due
to the blend with the strong C$_6$H line.}
\label{fig-cnminus-lineprofiles}
\end{figure}

To obtain a more reliable estimate of the abundance and excitation
conditions of CN$^-$ in IRC +10216 we have carried out radiative
transfer calculations based on the LVG formalism. The physical
parameters of the envelope have been taken from \citet{agu09}. We
included the first 20 rotational levels of CN$^-$. The rate
coefficients for de-excitation by collisions with para--H$_2$ have
been explicitly computed through quantum mechanical calculations
for temperatures up to 70 K and for transitions involving the
first 9 rotational levels of CN$^-$. The calculations are
described in the Appendix \ref{sec-collision-rates}. For
collisions with He, the rate coefficients computed for para--H$_2$
were scaled down by a factor of 1.37 (the ratio of the square
roots of the reduced mass of each couple of collision partners).
For transitions involving rotational levels higher than $J$ = 8
the Infinite Order Sudden approximation was used. As commented
before, CN$^-$ is confined to the outer envelope of IRC +10216. We
find that to reproduce the line profiles and relative intensities
observed, the abundance of CN$^-$ relative to H$_2$ must peak at a
radius between 13'' and 17'' from the star. The adopted radial
distribution, with a maximum abundance relative to H$_2$ of 2.5
$\times$ 10$^{-9}$ reached at a radius of 15'' (12'' if expressed
as particle density, see grey thick line in
Fig.~\ref{fig-chemistry}), produces line profiles in reasonable
agreement with the observed ones (see
Fig.~\ref{fig-cnminus-lineprofiles}). We note that since the
density decreases as the radius increases, the maximum in the
particle density is reached at smaller radii than the maximum in
the abundance relative to H$_2$. The total column density across
the envelope (twice the radial value) is 3 $\times$ 10$^{12}$
cm$^{-2}$, in good agreement with the value derived from the
rotational diagram. In the region where most of CN$^-$ is present
(at a radius of $\sim$ 2 $\times$ 10$^{16}$ cm, where the gas
kinetic temperature is $\sim$ 40 K and the density of H$_2$
molecules is around 4 $\times$ 10$^4$ cm$^{-3}$) the rotational
levels involved in the CN$^-$ observed transitions are
subthermally excited. Therefore, the collision rate coefficients
utilized turn out to be essential to correctly estimate the CN$^-$
abundance in the outer layers of IRC +10216's envelope.

To gain some insight into the formation of CN$^-$ in the external
layers of the molecular envelope of IRC +10216, we have performed
chemical modeling calculations similar to those described by
\citet{cer08}. The physical parameters of the envelope have been
taken from \citet{agu09}. The rate constants and branching ratios
of the reactions of anions with H, O, and N atoms, studied in the
laboratory by \citet{eic07}, have been updated according to the
values used by \citet{cor09} and
\citet{wal09}\footnote{\texttt{http://www.physics.ohio-state.edu/$\sim$eric/research.html}}.
Photodetachment rates of molecular anions have been assumed by
\citet{mil07} to depend on the electron affinity of the neutral
counterpart. For CN$^-$ we have assumed the same rate expression
adopted for C$_6$H$^-$, as the neutral counterparts of both
molecules have similar electron affinities (3.862 and 3.809 eV
respectively; \citealt{rie02}). Plotted in
Fig.~\ref{fig-chemistry} is the calculated radial distribution of
the abundance of CN$^-$ (black thin line) and of some other
molecular anions. CN$^-$ is predicted to form at a much greater
radius than C$_4$H$^-$, C$_6$H$^-$, C$_3$N$^-$, and C$_5$N$^-$,
because, unlike the other anions, it is not formed directly from
the radical CN but through the reactions of the anions C$_n^-$
($n$ = 5$-$10) with N atoms (see also \citealt{cor09}). CN being a
small molecule, the rate constant for the reaction of radiative
electron attachment is likely to be very small. Here we have
assumed a value of 2 $\times$ 10$^{-15}$ cm$^3$ s$^{-1}$ at 300 K,
similar to that computed for C$_2$H by \citet{her08}. This process
results in a too low formation rate for CN$^-$, more than 5 orders
of magnitude lower than that provided by the reactions of C$_n^-$
and N atoms. The reaction of HCN and H$^-$ is also a source of
CN$^-$ in the inner regions of the envelope, but has only a minor
contribution (less than 0.2 \%) to the total amount of CN$^-$
formed in the envelope. The anion C$_2$H$^-$, on the other hand,
is solely formed through the reaction of C$_2$H$_2$ and H$^-$,
which takes place in the inner regions. According to our chemical
model, CN$^-$ reaches a maximum abundance relative to H$_2$ of 1.6
$\times$ 10$^{-8}$ at a radius of 8 $\times$ 10$^{16}$ cm, and a
total column density across the envelope of 8 $\times$ 10$^{12}$
cm$^{-2}$. For C$_2$H$^-$, the model predicts a fairly small
column density of 7 $\times$ 10$^{10}$ cm$^{-2}$, distributed over
a region inner to 10$^{16}$ cm. These results are in agreement
with the recent chemical model of \citet{cor09}, who predicted
that both CN$^-$ and C$_2$H$^-$ could be detected in the
circumstellar envelope of IRC +10216.

The abundance and column density predicted for CN$^-$ by the
chemical model is in reasonable agreement with the value derived
from the observed lines and the LVG model. However, the calculated
spatial distribution is markedly different from that derived by
the observations (see Fig.~\ref{fig-chemistry}). In fact, by
adopting the CN$^-$ abundance distribution obtained with the
chemical model the resulting line profiles show important
discrepancies as compared with the observed ones. While the
calculated absolute line intensities are about the same order of
magnitude as those observed, significant disagreements between the
relative intensities and the line profiles are found. The
calculated line intensity decreases too rapidly when going from
the $J$ = 1-0 to the $J$ = 3-2, and the computed line profiles are
too much U-shaped, with nearly all the emission predicted to occur
at the line edges (i.e. at the terminal expansion velocity). These
discrepancies arise because the chemical model predicts that
CN$^-$ is present in a region of the circumstellar envelope that
is too far from the central star. An abundance distribution more
compact than predicted by our chemical model may arise if the
envelope is not modeled as being smooth, but as having
density-enhanced shells. \citet{cor09} have recently studied the
effect of such density enhancements on the radial distribution of
molecular abundances and found that molecules formed in the outer
envelope would concentrate at the position of the first and/or
second shells, located at 15 and 27'' respectively.

For C$_2$H$^-$, the chemical model predicts it to be distributed
over an 8'' diameter region (see Fig.~\ref{fig-chemistry}) with a
total column density of 7 $\times$ 10$^{10}$ cm$^{-2}$. Once
averaged over the 14.6'' beam of the IRAM 30-m telescope at the
frequency of the $J$ = 2-1 transition, the calculated column
density is about 3 times lower than, and thus consistent with, the
3$\sigma$ upper limit derived from the non detection of the $J$ =
2-1 line.

The identification of CN$^-$ in IRC +10216 with a relatively large
anion-to-neutral abundance ratio (0.25 \%) suggests that it may be
detectable in other astronomical sources. Upper limits to the
CN$^-$/CN abundance ratio as low as 0.2$-$2 \% were obtained in
TMC-1, L1527, Barnard 1, and the Orion Bar in a previous search
for the $J$ = 2-1 transition by \citet{agu08}. More sensitive
observations would be needed if the abundance of CN$^-$ in other
sources is similar to that found in IRC +10216.

The high abundance of CN$^-$ compared to that of C$_2$H$^-$
demonstrates the efficiency of the reactions of N atoms and large
carbon anions. A more sensitive search for C$_2$H$^-$ might
support this alternate scheme for the formation of anions in
space, and perhaps explain the low observed abundance of
C$_4$H$^-$ as compared to C$_3$N$^-$.

\begin{acknowledgements}

We acknowledge R. Chamberlin and T. G. Phillips for their kind
help during a previous search of the CN$^-$ $J$ = 3-2 transition
with the Caltech Submillimeter Observatory (CSO). We are also
grateful to the astronomers that helped with the observations
during the 2009 winter HERA pool at the IRAM 30-m telescope, among
them F. S. Tabatabaei, E. De Beck, G. Ba\~n\'o, and J. Rod\'on.
M.A. is supported by a \textit{Marie Curie Intra-European
Individual Fellowship} within the European Community 7th Framework
Programme under grant agreement n$^{\circ}$ 235753. J.R.G. is
supported by a Ram\'on y Cajal research contract from the Spanish
MICINN and co-financed by the European Social Fund. J.K.
acknowledges the partial financial supports from the University
Complutense of Madrid/Grupo Santander under the program of
Movilidad de Investigadores Extranjeros and from the U.S. National
Science Foundation under Grant No. CHE-0848110 to M. H. Alexander.
This project has been partly financed by the Spanish MICINN grants
Consolider-Ingenio 2010 CSD2009-00038, AYA2009-07304, and
CTQ2008-02578-BQU.

\end{acknowledgements}

\clearpage

\begin{appendix} 

\section{CN$^-$--H$_2$ collision rate coefficients}
\label{sec-collision-rates}

The potential energy surface (PES) of the CN$^-$--H$_{2}$ complex
was calculated {\em ab initio} using single and double-excitation
coupled cluster method with non-iterative triple excitations
[CCSD(T)] \citep{kno93,kno00} implemented in {\footnotesize
MOLPRO}\footnote{MOLPRO, version 2006.1, a package of ab initio
programs, H.-J. Werner, P. J. Knowles, R. Lindh, F. R. Manby, M.
Sch\"{u}tz, P. Celani, T. Korona, G. Rauhut, R. D. Amos, A.
Bernhardsson, A. Berning, D. L. Cooper, M. J. O. Deegan, A. J.
Dobbyn, F. Eckert, C. Hampel and G. Hetzer, A. W. Lloyd, S. J.
McNicholas, W. Meyer and M. E. Mura, A. Nicklass, P. Palmieri, R.
Pitzer, U. Schumann, H. Stoll, A. J. Stone, R. Tarroni and T.
Thorsteinsson, see \texttt{http://www.molpro.net}}. The geometry
of the system was described in the body-fixed frame and
characterized by three angles ($\theta$, $\theta'$, $\phi$) and
the distance $R$ between the centers of mass of H$_{2}$ and
CN$^-$. The H$_{2}$ bond distance was fixed at $r_{0}$=1.44876
a$_0$ and the CN$^{-}$ bond distance was varied for the purpose of
the averaging of the PES over the lowest vibrational state of the
CN$^-$ diatom. The basis set superposition error correction
counterpoise procedure of \citet{boy70} was applied. The four
atoms were described by the correlation-consistent triple zeta
basis set (aug-cc-pVTZ) of \citet{woo94} augmented by the (3s, 2p,
1d) midbond functions defined by \citet{wil95}, placed at
mid-distance between the CN$^-$ and H$_{2}$ centers of mass. The
final $V(r,R,\theta,\theta',\phi)$ PES is five-dimensional,
however, in this work we included only three perpendicular
orientations of the H$_2$ molecule [$(\theta',\phi)$ pairs:
$(0,0)$, $(0,90)$, $(90,90)$] to average over H$_2$ rotations.
Additionally, the PES was averaged over the CN$^-$ internuclear
distance corresponding to the CN$^-$ vibrational ground state wave
function. The 2-D PES was finally obtained as an arithmetic
average of three H$_2$ orientations. The full five-dimensional PES
and four-dimensional scattering calculations will be presented
elsewhere.

\begin{figure}
\includegraphics[angle=0,scale=0.475]{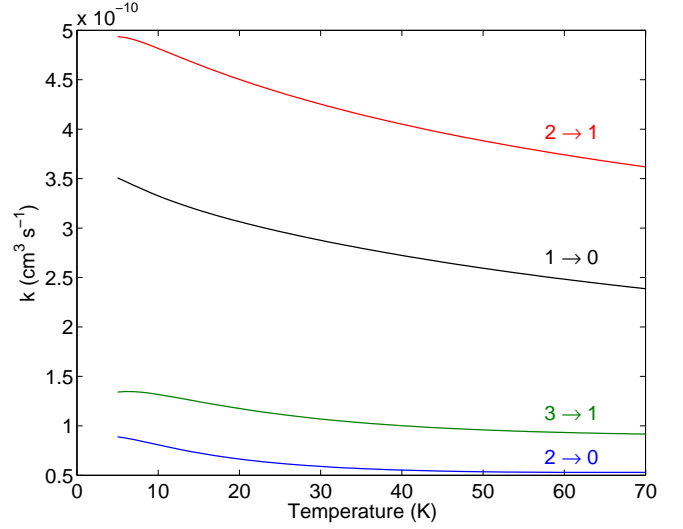}
\caption{Collisional de-excitation rate coefficients of CN$^-$ by
para--H$_2$ are shown as a function of temperature for the $J$ =
1-0, 2-1, 2-0, and 3-1 rotational transitions of CN$^-$.}
\label{fig-collision-rates}
\end{figure}

We considered collisions of CN$^{-}$ with para--H$_{2}(j_2=0)$ at
low temperatures. The rotational levels of CN$^-$ and H$_{2}$ are
designated by $j_{1}$ and $j_{2}$, respectively. We used the fully
quantum close-coupling approach of \citet{art60}. The standard
time-independent coupled scattering equations were solved using
the MOLSCAT code \citep{hut94}. Calculations were carried out at
values of the total energy ranging from 3.6 to 500 cm$^{-1}$. The
integration parameters were chosen to ensure convergence of the
cross sections over this range. At the largest total energy
considered  (500~cm$^{-1}$) the CN$^{-}$ rotational basis included
channels up to $j_{1}=21$ to ensure convergence of the excitation
functions $\sigma_{j_{1}j_{2} \to j_{1}'j_{2}'} (E_{c})$ for
transitions including up to the $j_{1}=8$ rotational level of
CN$^{-}$. The rotational basis of H$_{2}$ was restricted to
$j_2=0$ levels. The coupling with the $j_2=2$ (and higher) states
of H$_{2}$ was not taken into account. As shown by \citet{liq08},
this approach is expected to yield reliable results for the energy
range considered here. From the above described excitation
functions one can obtain the corresponding state-resolved thermal
rate coefficients by Boltzmann averaging:
\begin{eqnarray}
\label{thermal_average}
k_{j_{1}j_{2} \to j_{1}'j_{2}'}(T) & = & \left(\frac{8}{\pi\mu k^3 T^3}\right)^{1/2}  \nonumber\\
&  & \times  \int_{0}^{\infty} \sigma_{j_{1}j_{2} \to
j_{1}'j_{2}'}\, E_{c}\, e^{-E_c/kT}\, dE_{c}
\end{eqnarray}
where $k$ is the Boltzmann constant. To obtain precise values of
the rate constants, the energy grid was chosen to be sufficiently
fine to include the numerous scattering resonances. The total
energy range considered in this work allows us to determine rate
coefficients up to 70 K. The temperature dependence of the rate
coefficients for selected de-excitation transitions is illustrated
in Fig.~\ref{fig-collision-rates}, with the values given in
Table~\ref{table-collision-rates}.

\begin{table}
\caption{CN$^-$--H$_2$ collision rate coefficients ($10^{-10}$
cm$^3$ s$^{-1}$)} \label{table-collision-rates} \centering
\begin{tabular}{cccccccc}
\hline \hline
           & \multicolumn{7}{c}{Temperature (K)} \\
\cline{2-8}
Transition & 10   & 20   & 30   & 40   & 50   & 60   & 70 \\
\hline
1 $\to$ 0  & 3.33 & 3.06 & 2.88 & 2.72 & 2.59 & 2.48 & 2.39 \\
2 $\to$ 0  & 0.81 & 0.65 & 0.59 & 0.55 & 0.54 & 0.53 & 0.53 \\
2 $\to$ 1  & 4.81 & 4.50 & 4.25 & 4.05 & 3.88 & 3.74 & 3.62 \\
3 $\to$ 0  & 0.67 & 0.64 & 0.61 & 0.59 & 0.57 & 0.56 & 0.52 \\
3 $\to$ 1  & 1.32 & 1.18 & 1.07 & 1.00 & 0.96 & 0.93 & 0.92 \\
3 $\to$ 2  & 4.81 & 4.62 & 4.39 & 4.19 & 4.03 & 3.88 & 3.77 \\
\hline
\end{tabular}
\tablenotea{The complete set of de-excitation rate coefficients of
CN$^-$ by collisions with para--H$_2$ considered in this study is
available on the {\footnotesize BASECOL} website
\texttt{http://basecol.obspm.fr/}}
\end{table}

\end{appendix}

\end{document}